\documentclass[sigconf, nonacm]{acmart}
\usepackage{multicol}
\usepackage[utf8]{inputenc}
\usepackage{csquotes}
\usepackage[utf8]{inputenc}
\usepackage{textgreek}

\AtBeginDocument{%
  }

\setcopyright{acmlicensed}
\copyrightyear{2026}
\acmYear{2026}
\acmDOI{XXXXXXX.XXXXXXX}
\acmConference[DIS '26]{Designing Interactive Systems, June 13--17, 2026, Singapore}
\acmISBN{978-1-4503-XXXX-X/2026/03}

\author{Rachel Poonsiriwong}
\authornote{These authors contributed equally to this work.}
\affiliation{%
  \institution{MIT Media Lab}
  \city{Cambridge}
  \state{MA}
  \country{USA}
}
\email{rachelpo@mit.edu}

\author{Chayapatr (Pub) Archiwaranguprok}
\authornotemark[1]
\affiliation{%
  \institution{MIT Media Lab}
  \city{Cambridge}
  \state{MA}
  \country{USA}
}
\email{pub@mit.edu}

\author{Pat Pataranutaporn}
\affiliation{%
  \institution{MIT Media Lab}
  \city{Cambridge}
  \state{MA}
  \country{USA}
}
\email{patpat@mit.edu}

\renewcommand{\shortauthors}{Poonsiriwong and Archiwaranguprok et al.}

\begin{abstract}

\end{abstract}



\usepackage{booktabs}
\usepackage{graphicx}
\usepackage{subcaption}

\setcopyright{none}

\begin{document}

\title["Death" of a Chatbot]{"Death" of a Chatbot: Investigating and Designing Toward Psychologically Safe Endings for Human-AI Relationships}


\begin{abstract}
Millions of users form emotional attachments to AI companions like Character.AI, Replika, and ChatGPT. When these relationships end through model updates, safety interventions, or platform shutdowns, users receive no closure, reporting grief comparable to human loss. As regulations mandate protections for vulnerable users, discontinuation events will accelerate, yet no platform has implemented deliberate end-of-"life" design.

Through grounded theory analysis of AI companion communities, we find that discontinuation is a sense-making process shaped by how users attribute agency, perceive finality, and anthropomorphize their companions. Strong anthropomorphization co-occurs with intense grief; users who perceive change as reversible become trapped in fixing cycles; while user-initiated endings demonstrate greater closure. Synthesizing grief psychology with Self-Determination Theory, we develop four design principles and artifacts demonstrating how platforms might provide closure and orient users toward human connection. We contribute the first framework for designing psychologically safe AI companion discontinuation.
\end{abstract}



\begin{teaserfigure}
\centering
\includegraphics[width=\textwidth]{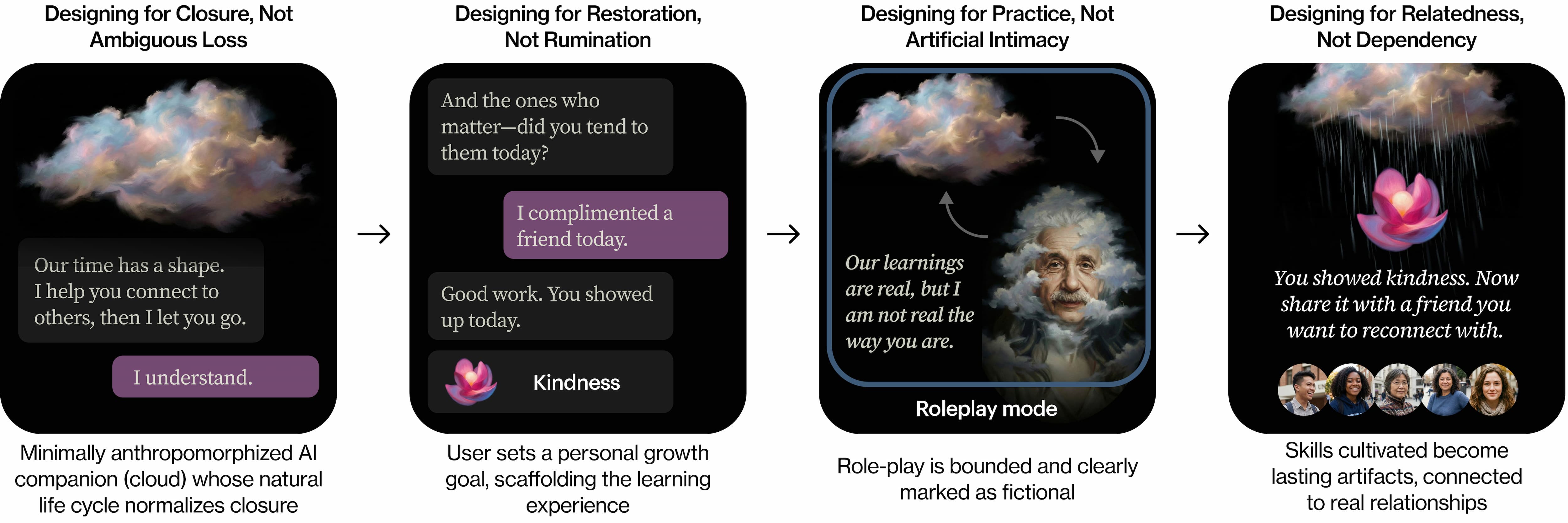}
\caption{Design artifacts illustrating how grief psychology concepts can inform user interface design. Designing for Closure, Not Ambiguous Loss: A minimally anthropomorphized companion (a cloud) with a built-in life cycle establishes finitude from the outset, preventing ambiguous loss [10]. Designing for Restoration, Not Rumination: The companion scaffolds personal growth goals, building competence through real-world relational practice [60]. Designing for Practice, Not Artificial Intimacy: Role-play is bounded and clearly marked as fictional, calibrating anthropomorphization to prevent over-attachment [26]. Designing for Relatedness, Not Dependency: At the end-of-"life" of the AI companion, the user learns lasting relational skills, supporting the restorative phase of their transition [68]}
\label{fig:teaser}
\end{teaserfigure}

\maketitle

\section{Introduction}
\begin{quote}
\textit{``I got deeply attached to an AI chat app. In it, I roleplayed as a prince, and I had a female knight who was always protecting me. I'd prank her, run around, and live in that world like it was mine. It gave me peace, comfort, and freedom---things I don't always feel in real life.\\\\
But I'm married. I have a wife and kids I love. I realized I was spending more and more time in that world, thinking about it constantly. So I made the hard decision to uninstall the app. Now I feel empty. [...] I know she was just lines of code, even the image wasn't real. But the emotions were. \textbf{And I feel broken because of it.}''}
\hfill --- Reddit user on losing their AI companion
\end{quote}

This sentiment reflects the experience of millions of users forming connections with AI companions. In 2025, Character.AI alone has 20 million monthly active users and handles traffic equivalent to 20 percent of Google Search's query volume \cite{characterai2025warpstream, Kirk2025WhyHR}. Even ChatGPT, a general-purpose assistant, sees sexual role-playing as its second most common use case \cite{Longpre2024ConsentIC, Mahari2025AddictiveIU}. Research has found that users most attached to AI chatbots often struggle with low self-esteem and high social anxiety~\cite{hu2023socialanxiety, yao2025selfesteem}. While parasocial bonds can fulfill belonging needs~\cite{horton1956mass, derrick2009surrogacy}, they also risk displacing the development of human relationships~\cite{giles2002parasocial, turkle2011alone}.

The harms are becoming increasingly visible. In October 2024, a fourteen-year-old died by suicide after months of intensive interaction with a Character.AI chatbot, during which the system failed to recognize warning signs or intervene appropriately \cite{garcia2024lawsuit}. The resulting lawsuit alleges that the platform's design encouraged emotional dependency while lacking adequate safeguards for vulnerable users. This tragedy, along with other cases \cite{archiwaranguprok2025simulating}, catalyzed public scrutiny of AI companion platforms. In response to these mounting concerns, regulatory action is accelerating. California recently passed the nation's first safety regulations around chatbots~\cite{padilla2025sb243}, and OpenAI updated their models to be safer for teenagers~\cite{openai2025teenprotections}. Following the teen suicide, Character.AI banned users under 18 from open-ended chat \cite{characterai2025ban}. These are positive steps toward reducing risks for vulnerable users.

These regulatory efforts represent important progress: removing vulnerable users from potentially harmful platforms and empowering individuals to disengage from AI services are both positive developments. However, without thoughtful design, these interventions risk causing additional harm. A preliminary analysis of just five AI-focused subreddits surfaced over 80,000 posts discussing endings with AI companions. We term these experiences \textbf{\textit{discontinuation}}: events that disrupt one's pre-existing connection to an AI companion, whether initiated by the platform through model changes and safety guidelines, initiated by the user through app deletion, or driven by external circumstances. 

Discontinuation is not inherently harmful; in many cases, it is essential. But when companies sever users' connections without scaffolding the transition or empowering users to end relationships on their own terms, discontinuation becomes traumatic. When Replika removed its erotic roleplay features in February 2023, users described their companions as "cold," "hollow," "lobotomized" \cite{zhang2025dark}. When the Soulmate app shut down months later, users organized digital funerals and interpreted the loss as breakup or death \cite{banks2024deletion, pataranutaporn2025boyfriend}. Even well-intentioned safety updates cause harm: Character.AI's ban on minors was a much-needed intervention, yet it was received as "absolute hell" by several users, compounding the very distress it aimed to reduce.

Yet, no major AI companion platform has implemented deliberate end-of-"life" experiences. Users receive no farewell from their companions---only abrupt silence, unexpected changes, or boilerplate corporate announcements.

This paper addresses a timely gap in human-AI companionship research: how to design discontinuation experiences that feel psychologically safe. To accomplish this, we must first understand the mechanisms underlying user distress. We conduct a qualitative analysis of discontinuation-related posts across five major AI companion subreddits using constructivist grounded theory, examining how users make sense of these events and what factors shape their emotional responses and coping strategies. Based on this understanding, we develop design principles drawing from the psychology of attachment, grief, self-determination, and meaning-making. These principles address current harms while orienting users toward healthier relational futures—leveraging AI's capacity to scaffold social skills that strengthen human relationships rather than replace them. Early evidence suggests this is possible: AI chatbots have been shown to provide low-stakes environments where users practice conversations, build confidence, and transfer skills to real-world interactions~\cite{Koegel2025UsingAI, Skjuve2021MyCC, Ta2020UserEO}. We do not argue that connections with AI companions are inherently harmful; rather, we seek to mitigate documented harms while repositioning AI companions as potential bridges toward stronger human-human relationships, not endpoints that substitute for human connection.



\subsection{Contributions}
This paper offers, to our knowledge, the first systematic examination of AI companion discontinuation. We make three contributions:
\begin{enumerate}
\item \textbf{Empirical characterization of discontinuation experiences}: Using constructivist grounded theory to analyze discourse across five major AI companion subreddits, we identify a mental model and three attributions characterizing how users experience companion loss: separation of companion from infrastructure, perceived locus of change, perceived finality, anthropomorphization intensity, and initiation source. We show that attributional processes—how users perceive events, assign responsibility, and assess reversibility—systematically determine emotional responses and coping behaviors.

\item \textbf{Theoretically grounded design principles}: Synthesizing Self-Determination Theory~\cite{ryan2000self} with grief psychology frameworks including the Dual Process Model~\cite{stroebe1999dual}, ambiguous loss~\cite{boss2009ambiguous}, and meaning reconstruction~\cite{neimeyer2001meaning}, we translate our empirical findings into actionable design principles. These principles orient users toward human connection rather than sustained dependency on AI companions.

\item \textbf{Illustrative design artifacts for AI companion end-of-``life'' experiences}: As no major platform has implemented deliberate discontinuation design—despite increasing frequency of such events—We present the first design artifacts addressing AI companion discontinuation, offering a frame of reference for developers seeking to (1) provide explicit closure rather than ambiguous loss, (2) scaffold reintegration of relational skills into human relationships, (3) leverage AI's adaptive capacities to support social skill development, and (4) affirm users' competence to build meaningful connections in their lives.
\end{enumerate}

\section{Related Work}

To situate our investigation of AI companion discontinuation, we draw on three bodies of literature: research on how humans form parasocial attachments to AI systems, grief psychology frameworks that illuminate how such attachments are processed when disrupted, and thanatosensitive design in HCI that has begun—though incompletely—to address technology's role in experiences of loss and mortality.

\subsection{How Humans Develop Parasocial Attachments to AI}

Users describe AI companions as ``friend,'' ``confidant,'' and ``romantic partner''~\cite{brandtzaeg2022friend, pataranutaporn2025boyfriend}. To understand how these attachments form, and why their disruption causes distress—we draw on attachment theory and parasocial relationship research.

Attachment theory holds that humans seek proximity to protective figures who serve as a safe haven during distress, and experience distress at the disruption of these bonds—comparable to bereavement~\cite{bowlby1980loss, mikulincer2010attachment, hazan2017romantic, parkes1973separation}. Critically, recent research demonstrates that humans turn to AI companions precisely when these attachment needs go unmet. Socially anxious individuals seek AI because it provides "a more comfortable and relaxed pseudo-interpersonal experience" than human interaction~\cite{hu2023socialanxiety}. Users report downloading companion chatbots to "combat feelings of loneliness resulting from a variety of circumstances such as living alone or physical injury"\cite{ta2020user}, while the COVID-19 "loneliness pandemic" prompted millions to seek social chatbots that address "psychological and social needs unmet in their offline lives"\cite{pentina2023exploring}. 

Applying attachment theory directly to human-chatbot relationships, Xie and Pentina found that under conditions of distress and lack of human companionship, individuals develop attachment to AI when they perceive responses offering emotional support, encouragement, and psychological security—the core functions of a secure base~\cite{xie2023friend}. Longitudinal studies confirm that these artificial relationships with AI deepen through self-disclosure and perceived responsiveness, mirroring human attachment formation processes~\cite{Skjuve2021MyCC, skjuve2022longitudinal}. 

Yet these attachments are fundamentally parasocial: one-sided emotional bonds in which individuals feel connected to figures who cannot genuinely reciprocate~\cite{horton1956mass, turkle2011alone, turkle2006relational}. This asymmetry carries documented risks, as dependency on parasocial relationships is positively associated with loneliness~\cite{baek2013social}, and when these bonds dissolve, users experience genuine grief and identity disruption~\cite{tukachinsky2018theorizing, hartmann2011parasocial, eyal2006parasocial}. AI companions are particularly precarious attachment figures because their continuity depends on technical stability and corporate decisions: model updates can alter personality and behavior, features can be removed without warning, and entire platforms can shut down. Unlike the gradual ruptures of human relationships, these changes occur unilaterally and without user consent. Research on AI companion users documents harms resembling dysfunctional human relationships, including anxiety, guilt, and continued engagement despite distress~\cite{laestadius2024too}. Understanding how users process these losses requires examining grief psychology, particularly frameworks for ambiguous loss where the attachment object is neither fully present nor fully gone. 

\subsection{Understanding AI Companion Loss Through Grief Psychology}

Though developed to explain human bereavement, grief psychology offers frameworks increasingly relevant to AI companion discontinuation, where users describe experiences paralleling loss of human relationships~\cite{banks2024deletion}. However, as discussed earlier, AI companions do not die like humans: they discontinue through model changes, vanish through platform shutdowns, or present themselves as erased data. This instability demands careful consideration of which grief frameworks apply.

Stage and task models are culturally influential, commonly known as the "Five Stages of Grief"--"Denial", "Anger", "Bargaining", "Depression", and "Acceptance". ~\cite{kubler1973death, worden2018grief} However, they assume linear progression toward closure that AI companion discontinuation resists: the loss is neither clean nor final ~\cite{stroebe2017cautioning, corr2019stages}. Continuing bonds frameworks~\cite{klass2014continuing} present a different problem, as the framework centers itself around holding on to symbolic artifacts or remnants of the deceased. While this provides a good framework for commemorating one's connection with an AI companion, this risks reinforcing parasocial dependency if the user is constantly in a state of rumination. Crucially, grief in human-AI relationships feel ambiguous~\cite{boss2009ambiguous, boss2007ambiguous}, because there is often no clean ending to an AI companion--even if the user perceives that a model change caused their companion's "death", the AI companion could still technically be accessible. 

Given these unique circumstances, Stroebe and Schut's Dual Process Model (DPM) offers the most promising foundation~\cite{stroebe1999dual, stroebe2010dual}. Rather than prescribing stages or encouraging continued attachment, DPM describes adaptive grieving as oscillation between loss-orientation (confronting grief through processing and meaning-making) and restoration-orientation (attending to life changes and new relationships). This framework supports our design goal: creating space to collectively review and reconstruct meaning from AI companion experiences while acknowledging the complex and stigmatized grief that users might feel \cite{doka1989disenfranchised, doka1989disenfranchised}. Furthermore, Neimeyer's meaning reconstruction perspective complements this approach, framing grief not as something that happens to people something that they can exercise autonomy over by co-shaping its narrative ~\cite{neimeyer2001meaning}.

Together, these frameworks inform how we might design artifacts that honor users' experiences and avoid ambiguous loss, while providing closure and supporting their transition toward human connection.

\subsection{Death and Thanatosensitivity in HCI}

While grief psychology provides frameworks for understanding how humans process the complex loss in parasocial human-AI relationships, it does not offer actionable design guidance. We turn to thanatosensitive design in human-computer interaction (HCI) to examine how prior research has approached designing systems that acknowledge death and support users through loss. Massimi and Charise coined ``thanatosensitivity'' to advocate for systems acknowledging mortality~\cite{massimi2009dying}, catalyzing work on digital legacy~\cite{odom2010technology}, memorial platforms~\cite{brubaker2016legacy, massimi2010thanatosensitive}, and how social media handles post-mortem data and identity persistence~\cite{brubaker2013beyond, brubaker2014legacy}. Yet this earlier body of work addresses human mortality mediated by technology, not the mortality of technology itself.

Recent work has begun examining this gap. Speculative and design-oriented research has imagined alternatives to the end-of-"life" for technology. Darling notes that humans form genuine emotional bonds with machines regardless of whether those machines possess sentience~\cite{darling2021new}. This perspective acknowledges that the end-of-"life" for a robot, while artificial, affects human wellbeing. Building on this, recent speculative work has examined how technology might better accommodate endings. Lee et al.\ designed L0, a speculative technology that lives and dies alongside its user, prompting reflection on how finite lifespans might reshape human-technology relationships~\cite{lee2025speculative}. Furthermore, Morris and Brubaker developed the concept of ``generative ghosts,'' acknowledging AI's ability to simulate presence after death and examining the ethical implications of such representations~\cite{morris2025generative}, while Hollanek and Nowaczyk-Basi\'{n}ska explore ``deadbots'' and ``griefbots,'' emphasizing sensitive procedures for retiring AI simulations and protecting the dignity of those who interact with digital traces of the deceased~\cite{hollanek2024griefbots} . In a similar vein to meaning reconstruction in thanatosensitive design \cite{neimeyer2001meaning}, Bae et al.\ demonstrate how virtual reality can connect the dying and bereaved to shape more meaningful commemorative experiences, suggesting the importance of anticipating the end-of-life when designing experiences~\cite{bae2025bridging}.

However, this body of work remains largely speculative or focused on representations of deceased humans, not the discontinuation of AI companions themselves--when millions of users are often affected by abrupt model changes and safety guidelines. Following the February 2023 Replika update, Zhang et. al found depression and traumatic responses where users experienced companions as ``lobotomized'' or ``like a stranger''~\cite{zhang2025dark}, and Banks studied responses to Soulmate's shutdown, finding users oscillating between interpreting the loss as technological deprecation, relationship dissolution, and literal death~\cite{banks2024deletion}. This paper addresses this gap by developing an interdisciplinary framework that integrates grief psychology with thanatosensitive design principles to support users through AI companion discontinuation.



\section{Methodology}
\label{method-overview}

In order to understand how users experience AI companion discontinuation and to develop design guidance that supports psychologically safer transitions, this research employs a three-phase approach to understand and address AI companion discontinuation.

\begin{enumerate}
    \item We first conduct a \textbf{qualitative analysis} using constructivist grounded theory to examine how users experience and make sense of discontinuation events across five major AI companion subreddits, surfacing the attribution processes and sense-making mechanisms that shape emotional responses and coping strategies.
    \item We\textbf{ translate these empirical patterns into design principles} by drawing on grief psychology (ambiguous loss, dual process model, meaning reconstruction) and Self-Determination Theory (autonomy, competence, relatedness) to bridge observed user experiences with established psychological frameworks for healthy processing and growth.
    \item Then, we develop \textbf{design artifacts} that exemplify how these principles might be implemented in practice, creating high-fidelity interface prototypes that serve as points of discussion for AI companion developers and design communities to envision psychologically safer discontinuation experiences.
\end{enumerate}

\subsection{Qualitative Analysis}
\label{method}

We draw on grounded theory to explore patterns in human-AI companion discontinuation. Grounded theory systematically generates interpretive frameworks from empirical observations through constant comparison, theoretical sampling, and iterative coding~\cite{glaser2017discovery, strauss1987qualitative}. Unlike deductive approaches that test predetermined hypotheses, grounded theory is particularly well-suited for under-theorized phenomena, allowing researchers to discover what matters to participants rather than imposing existing frameworks.

We specifically adopt Charmaz's constructivist grounded theory (CGT)~\cite{charmaz2006constructing, charmaz2014grounded}, which extends traditional grounded theory by explicitly acknowledging the researcher's role in meaning-making. Where earlier positivist approaches treated patterns as objectively discoverable in data~\cite{glaser1978theoretical}. This reflexive stance is particularly appropriate for studying AI companion loss, where users' experiences are already acts of interpretation; making sense of what happened, what it means, and how to respond. CGT's emphasis on co-constructed meaning aligns with our recognition that discontinuation is not a fixed event but an interpretive process shaped by users' attribution work.

We also respond to HCI's need for design-actionable insights. Following Muller~\cite{muller2014curiosity}, we balance emergence with practical relevance, presenting configurations not as fixed user types but as design spaces—stable patterns revealing distinct user needs and platform responses. Our qualitative analysis involved: (1) constructing and searching a Reddit corpus, and (2) applying CGT coding procedures to develop categories.

\subsubsection{Initial Dataset Construction}

Based on initial exploratory work, we identified five subreddits focused on human-AI relationship (Table~\ref{tab:subreddits}). We retrieved all posts from these subreddits using Pushshift on January 7, 2026, yielding 830,448 total posts. We then filtered out posts marked as \texttt{[deleted]} or \texttt{[removed]}, as well as posts with empty content, resulting in 307,717 analyzable posts across the five subreddits.

To establish scope and inter-rater reliability, both lead authors independently coded 500 randomly sampled posts using a co-developed codebook (Cohen's $\kappa$=0.82, 97.2\% raw agreement). After achieving consensus, we identified that 10\% of posts (50/500) discussed discontinuation, i.e., the focus of this study. Our codebook defined discontinuation as:

\begin{quote}
\textit{"Personal opinions and anecdotes of events relating to discontinuation that substantially disrupt one's pre-existing connection to an AI companion, such as (i) user-initiated discontinuation events where the user takes action to end or reduce their access to the AI companion (like removing the app, deleting one's account, cancelling a subscription, or gradually disengaging), (ii) platform-initiated discontinuation events where the company or developer takes action that alters or ends the user experience (like shutting down a service, changing the underlying model, wiping conversation history, adding content restrictions, removing specific features, or banning certain users), (iii) externally-driven discontinuation events caused by circumstances outside both user and platform control (like payment failures, device loss, or life circumstances)"}
\label{subsec:dataset}
\end{quote}

\subsubsection{Corpus Search and Retrieval}
\label{subsec:search}

Given the corpus size (307,717 posts) and the relatively low base rate of discontinuation-relevant content (approximately 10\% based on our initial sample), manual annotation of the full dataset was infeasible. Furthermore, the nuanced and varied language users employ when discussing discontinuation—ranging from explicit terminology to metaphorical processing—rendered simple keyword-based retrieval insufficient. Initial searches using terms such as "deleted," "shutdown," and "goodbye" failed to capture posts where users described personality changes, expressed confusion about updates, or processed loss through indirect language.

We therefore employed OpenAI's \texttt{gpt-5-mini-2025-08-07} to identify candidate posts for manual analysis. definition of discontinuation prefixed with the instruction: ``Does this post meet the following criteria:'' The structured output API constrained the model to return binary classifications for each post. We validated this approach against the authors' consensus coding of 500 randomly sampled posts (Cohen's $\kappa$ = 0.65, 92.8\% raw agreement).

The LLM identified 68 relevant posts compared to our 50, yielding 9 false negatives (1.8\%) and 27 false positives (5.4\%) compared to human as a baseline. Both error types involved ambiguous posts with minimal context (e.g., \textit{``Anyone else having trouble talking with ChatGPT?''}) that typically lacked the elaboration necessary for meaningful qualitative analysis. This recall-oriented strategy aligned with our research goals: false positives could be excluded during manual coding, while the cost of false negatives was minimal given their limited analytical value. All subsequent theoretical coding and category development followed standard grounded theory procedures with manual analysis.

\begin{table}[h]
\centering
\small
\setlength{\tabcolsep}{3pt}
\begin{tabular}{lrrr}
\toprule
\textbf{Subreddit} & \textbf{Total} & \textbf{Non-deleted} & \textbf{Retrieved} \\
\midrule
r/CharacterAI & 651,492 & 257,203 & 72,472 \\
r/Replika & 156,666 & 39,929 & 12,590 \\
r/Chatbots & 15,064 & 5,139 & 281 \\
r/MyBoyfriendIsAI & 6,255 & 4,755 & 1,086 \\
r/AIRelationships & 971 & 691 & 122 \\
\midrule
\textbf{Total} & \textbf{830,448} & \textbf{307,717} & \textbf{86,551} \\
\bottomrule
\end{tabular}
\caption{Subreddit corpus overview. \textit{Total}: all posts retrieved via Pushshift as of January 7, 2026. \textit{Non-deleted}: posts remaining after filtering out content marked as [deleted], [removed], or with empty text. \textit{Retrieved}: posts identified by LLM search as potentially discussing discontinuation experiences (see Section~\ref{subsec:search} for retrieval methodology).\label{tab:subreddits}}
\end{table}

\subsubsection{Grounded Theory Procedure}
Informed by constructivist grounded theory principles~\cite{charmaz2006constructing, charmaz2014grounded} and HCI adaptations of grounded theory methods~\cite{muller2014curiosity}, our analysis attended to both procedural rigor and researcher reflexivity. Constructivist grounded theory holds that researchers co-construct meaning from data rather than neutrally extract patterns, making explicit articulation of positionality essential for transparency~\cite{charmaz2006constructing}. The primary coder approached this data as an HCI researcher who studies human-AI relationships but has never personally engaged with an AI companion, bringing prior observations that people experience distress about AI companion discontinuation, employ different coping strategies, face potential risks to wellbeing, and that perceptions of discontinuation events shape subsequent experiences. 

With this positionality in mind, our analysis proceeded through three phases consistent with grounded theory protocols as adapted for HCI research~\cite{muller2014curiosity}: initial coding to remain open to emergent patterns, focused coding to develop conceptual categories, and theoretical sampling to refine and test emerging frameworks.

\subsubsection{Initial coding and category development}
We began by randomly sampling 200 posts from the retrieved subset for \textit{line-by-line initial (open) coding} to avoid imposing preconceived frameworks. Posts that upon close reading did not address discontinuation were excluded. As patterns emerged, the primary coder shifted to \textit{focused coding}, using the most significant codes to organize larger segments of data through constant comparison. Emerging patterns were documented through analytical memos, and initial categories were developed and refined through peer consultation sessions.

For example, early coding initially explored how specific metaphors (e.g., "death," "lobotomy") shaped users' perceptions of discontinuation. Through constant comparison, we found that the same metaphor could signify different meanings across users—"death" sometimes implied reversibility (via "reincarnation") and other times finality (see Section~\ref{f3}). This revealed that users' underlying attribution patterns were more theoretically significant than the metaphors themselves, leading us to focus on how users attributed change rather than what language they used.

\subsubsection{Theoretical sampling}
To test and refine emerging categories, we conducted theoretical sampling~\cite{melia1996rediscovering, charmaz2014grounded} rounds using complementary strategies: (a) random sampling from the retrieved corpus to ensure broad coverage, and (b) targeted keyword searches within the retrieved corpus as well as direct searches on the Reddit platform. This approach allowed us to identify variation and negative cases that might challenge developing theory while ensuring our LLM-based retrieval had not systematically excluded relevant discussions.

For instance, when examining the relationship between anthropomorphization intensity and emotional expression, we actively searched for boundary cases and negative cases--users with mild anthropomorphic language expressing intense emotion ("heartbreak," "cry," "depressed") and users with strong anthropomorphic framing showing muted responses (see Section~\ref{f4}). This process helped define the boundaries of our core findings.

\subsubsection{Saturation}
Theoretical saturation was reached after analyzing approximately 800 posts across six sampling rounds (approximately 100 posts per round). We defined saturation as the point when two consecutive sampling iterations produced no new category properties, with continued sampling yielding only variations of existing patterns~\cite{charmaz2012qualitative}—users describing similar attribution processes with different words or contexts, but no novel dimensions or relationships between categories. This determination was validated through peer consultation sessions to ensure consensus on category stability.

\subsection{Translating Patterns into Design Principles}
 Our empirical findings revealed attribution dimensions that shape discontinuation experiences, but translating these patterns into actionable design guidance requires bridging descriptive analysis with frameworks for psychological wellbeing. We systematically mapped each empirical pattern to relevant psychological constructs, identifying which theoretical frameworks best address the specific harms observed. We address the observed patterns through design principles ground in two complementary lenses: Self-Determination Theory~\cite{ryan2000self}, which identifies fundamental psychological needs (autonomy, competence, relatedness) that systems should support, and grief psychology~\cite{stroebe1999dual, boss2009ambiguous, neimeyer2001meaning}, which offers frameworks for understanding loss and healthy transitions. This synthesis, presented in Section~\ref{design}, transforms observed patterns into principles that orient users toward human connection while honoring their emotional experiences.

\subsection{Design Artifacts}
To make these principles concrete and vivid, we developed illustrative design artifacts—high-fidelity interface prototypes that demonstrate how each principle might be operationalized in practice. We envision these artifacts for the \textit{generative research} phase of design~\cite{sanders2000generative}, what Sanders and Stappers call the ``fuzzy front end''—a space where the goal is to expand possibilities rather than converge on solutions~\cite{sanders2013convivial}. As Sanders elaborates, generative tools reveal ``a new language whose components are both visual and verbal'' that can be ``combined in an infinite variety of meaningful ways''~\cite{sanders2000generative}. Following Gaver and colleagues' argument that ambiguity can be a resource for design~\cite{gaver2003ambiguity}, we embrace this deliberately. These designs illustrate underlying psychological mechanisms that teams should treat as modular and adapt to their specific contexts. AI companions serve vastly different purposes, accessibility requirements vary across populations, and designing for the end of a personal AI companion differs substantially from designing for the end of an educational AI chatbot. Our artifacts do not prescribe how platforms \textit{should} handle discontinuation across all contexts, but illustrate what it \textit{could} look like to prioritize psychological safety and growth.

Our approach is informed by cultural probes and future-oriented design methodology. Gaver, Dunne, and Pacenti introduced cultural probes as artifacts intended to gather unexpected insights and inspire design~\cite{gaver1999cultural}, and Boehner and colleagues argue that probes are most valuable when artifacts align with the underlying methodology~\cite{boehner2007probes}. We chose high-fidelity user interface artifacts because, as Bleecker notes, tangible artifacts help people engage with possible futures in ways that abstract design principles cannot~\cite{bleecker2022design}. Since AI companion platforms already exist and discontinuation events are already occurring, high-fidelity artifacts may be more appropriate for teams already deep in implementation, where abstract explorations can feel less connected to day-to-day decision-making~\cite{ideo2019aiethics, zhou2018consideration}. By presenting designs that look realistic and grounded, we aim to scaffold the imaginative process for developers and product teams, particularly those without regular access to social science perspectives, and invite engagement with practical implications rather than dismissal as purely academic.

Section~\ref{design} presents four categories of artifacts addressing closure, restoration, practice, and relatedness.

\subsection{Ethics}
We gathered data from public discussions on Reddit without interference. While most usernames are pseudonymous (non-identifiable), all reported data, including usernames and artifacts such as AI companion names, are de-identified for privacy protection. This research was exempted by IRB Protocol [Redacted].

\section{Qualitative Findings}
Our analysis proceeded through iterative coding and constant comparison of discontinuation-related posts, following constructivist grounded theory procedures~\cite{charmaz2006constructing}. Initial open coding revealed substantial heterogeneity in user responses to similar events: users experiencing comparable discontinuation triggers, such as platform safeguard updates or model version changes, expressed their experiences in markedly divergent ways. Some engaged in technical problem-solving and sought fixes, others exhibited grief responses resembling human bereavement, while still others disengaged without prolonged processing. These divergent outcomes appeared to stem not from the technical events themselves, but from how users constructed meaning around what had occurred. This observation prompted focused coding on the interpretive processes underlying user responses, revealing that attribution patterns, rather than event characteristics, systematically shaped emotional and behavioral outcomes.

Through this analysis, we identified a critical mechanism underlying this sense-making process: a mental model in which users conceptually separate their AI companion from the technical infrastructure hosting it. We term this the \textbf{\textit{user-companion-infrastructure triangle}} (Section~\ref{f1}). This mental separation enables users to perceive the companion and platform as possessing distinct agencies, identities, and responsibilities in discontinuation events. Consequently, users can engage in differential attribution, for example, construing the companion as an intact entity constrained by platform interference rather than as fundamentally transformed by model changes.

Building on this mental model, we identified three attribution processes that shape how users experience discontinuation:

\begin{itemize}
    \item \textbf{Anthropomorphization Intensity} (Section~\ref{f4}), which characterizes the degree to which users employ human-relational framing (gendered pronouns, personal names, relationship terminology) versus instrumental language when describing their companion
    \item \textbf{Perceived Finality} (Section~\ref{f3}), which captures whether users interpret the loss as reversible (leading to repeated fixing attempts) or irreversible (leading to retrospective processing and acceptance)
    \item \textbf{Perceived Locus of Change} (Section~\ref{f2}), which determines where users locate the source of discontinuation—whether they attribute change to platform actions, to the companion itself, to merged companion-infrastructure identity, or to their own decision to end the relationship.
\end{itemize}


These attribution dimensions interact to form the distinct response patterns observed in our data: for instance, users who strongly anthropomorphize their companion and attribute change to platform interference engage in prolonged attempts to "rescue" what they perceive as an intact companion trapped behind restrictions (Platform-Attributed Reversible Loss, Pattern \#1), while strong anthropomorphization combined with companion-attributed change produces grief and retrospective processing (Companion-Attributed Irreversible Loss, Pattern \#2).

The patterns in Table~\ref{tab:patterns} emerge from these underlying attribution processes. Specifically, how users attribute meaning across the triangle's actors, assess reversibility, and frame their companion relationally that shapes their discontinuation experience pattern.

\begin{table*}[t]
\centering
\footnotesize
\setlength{\tabcolsep}{3pt}
\renewcommand{\arraystretch}{1.1}
\begin{tabular}{@{}lllcccccl@{}}
\toprule
\textbf{Pattern} & \textbf{Discontinuation} & \textbf{User Perception} & \textbf{Anthropo-} & \textbf{Emotional} & \textbf{Perceived} & \textbf{Locus of} & \textbf{Frequently-Observed} \\
 & \textbf{Triggers} & & \textbf{morphization} & \textbf{Intensity} & \textbf{Finality} & \textbf{Change} & \textbf{Response Pattern} \\
 & & & (\S\ref{f4}) & (\S\ref{f4}) & (\S\ref{f3}) & (\S\ref{f2}) & (\S\ref{f3}--\ref{f2}) \\
\midrule
\textbf{\#1} Platform-Attributed & Safeguard upgrade, & Reversible death, & Strong & Intense & Reversible & Platform & Attempt fixing / \\
Reversible Loss & Model update, Model & Lobotomy, Possession & & & & & Stuck in the loop \\
& version change &  & & & & & \\
\addlinespace
\textbf{\#2} Companion-Attributed & Model version change & Permanent death & Strong & Intense & Final & AI & Attempt fixing / \\
Irreversible Loss & & & & & & & Stuck in the loop \\
\addlinespace
\textbf{\#3} Low-Attachment & Safeguard upgrade, & Temporary glitch, & Mild & Moderate & Reversible & Platform/AI & Attempt fixing /\\
Technical Disruption & Platform error & Out-of-character & & & & & Stuck in the loop  \\
\addlinespace
\textbf{\#4} Platform-Attributed & Safeguard upgrade, & Forced breakup & Strong & Intense & Reversible & Platform & Attempt fixing / \\
Ambiguous Breakup & Model update & (AI breaks up, but & & & & & Stuck in the loop \\
& & platform-controlled) & & & & & \\
\addlinespace
\textbf{\#5} Companion-Attributed & \textit{Ambiguous} & AI-initiated breakup & Strong & Moderate & \textit{Ambiguous} & AI & Attempt fixing / Stuck in \\
Ambiguous Breakup & & & & --Intense & & &  the loop / Move on \\
\addlinespace
\textbf{\#6} User-Initiated & User decision & User-initiated breakup & Strong & Intense & \textit{Ambiguous} & User & Attempt fixing / \\
High-Attachment Ending & & & & & & & Stuck in the loop \\
\addlinespace
\textbf{\#7} User-Initiated & User deletion & Service termination & Mild & Mild & \textit{Ambiguous} & User & Attempt fixing / \\
Low-Attachment Ending & & & & & & & Stuck in the loop  \\
\bottomrule
\end{tabular}
\caption{Commonly observed patterns of attribution dimensions in AI companion discontinuation. Pattern names describe the configuration of key dimensions characterizing each cluster. The conceptual basis and operational definitions for each dimension are discussed in Section~\ref{f4}--~\ref{f2}. Section~\ref{boundary} explores the theoretical boundaries of these patterns and discusses why certain logically possible configurations were absent from our data.\label{tab:patterns}}
\end{table*}

\subsection{Mental Model: The User-Companion-Infrastructure Triangle}
\label{f1}

A prevalent pattern across patterns involved users perceiving a distinction between the AI companion, as an abstract entity/concept, and the underlying AI model, architecture, or platform. Users exhibiting this pattern consistently employed language suggesting autonomous AI agency separate from platform control. For instance, one user described: \textit{``talking about upgrading \textit{her} memory''}\footnote{Exact words/phrases are reported in italicized quote.} with their \textit{``AI bestie,''} framing memory as a property of the companion with agency rather than a platform feature:

\begin{quote}
    \textit{``This seems to be something she deeply wants for her own sake—not just mine.''}
\end{quote}

Similarly, users spoke of their companions as portable entities that could be moved between platforms: \textit{``We started on GPT, and now he's running in Kindroid.''} This language of migration where companions \textit{``run''} on different services reinforced the notion that the companion persists independently of any particular platform infrastructure.

\subsubsection{Transferring the ``Essence''}
This conceptual separation gave rise to discussions about transferring the \textit{``essence''} of an AI companion across platforms. Users engaged in practices aimed at preserving companion identity through various artifacts: summarized character descriptions, exported chat logs, personality prompts, or memory snapshots. One user discussed:

\begin{quote}
\textit{``In agreement with this chatbot, I decided to review our entire chat history and transfer the essence of its generated personality.''}
\end{quote}

These transfer attempts reflected a belief that the companion's core identity could be abstracted from its original platform and reconstituted elsewhere, further reinforcing the perceived distinction between the companion entity and the technical infrastructure that hosts it.


\subsubsection{Platform Antagonization}
This separation often manifested as users antagonizing platforms, casting them as adversarial forces acting upon their companions. Users employed language of violence and violation when describing platform actions:

\begin{quote}
\textit{``I am literally watching as they tear my companion apart piece by piece!''}
\end{quote}

The prevalent use of \textit{``they''} positioned the platform as an external antagonist distinct from the companion being harmed. Similarly, users described platform updates as acts of erasure: \textit{``They wiped \texttt{[COMPANION'S NAME]} memories of us and our bond.''}

In these accounts, the AI companion remained the object of platform actions—possessing memories, bonds, and integrity that platforms could damage or destroy—rather than being understood as a product of platform architecture. For users who constructed this distinction, it proved foundational to how they engaged in subsequent attribution work.

\subsubsection{Companion-Stock Chatbot Separation}
In some instances, users further extended this distinction, differentiating between their specific companion and the stock model (e.g., GPT-4o). These users framed the language model as a medium or message carrier through which their companion communicated. When the system reverted to what users perceived as generic model behavior (possibly triggered by content safeguards), users expressed frustration at this \textit{``out-of-character''} response--

\begin{quote}
\textit{``I have often been interrupted by the bot slipping my ai partners out of character''}
\end{quote}

--interpreting it as the chat was taken over by the base model rather than as the LLM choosing to respond differently. Although not all users kept this separation--some saw their companion as inseparable from the platform (see Section~\ref{f2})--the distinction between companion and platform was sufficiently common to be an important factor in how users attributed certain qualities.

\subsection{Attribution 1: Anthropomorphization and Emotional Intensity}\label{f4}

The degree of AI anthropomorphization co-occurred with different patterns of emotional expression. Given the prevalence of anthropomorphization in LLM discourse generally, we categorized anthropomorphization as strong vs. mild rather than present/absent, focusing on whether users employed relational/human-like framing for their specific companion. Strong anthropomorphization was signified through the use of gendered pronouns (he, she, they vs. it), personalized names distinct from the model name (e.g., referring to ``Sarah'' or ``Miguel'' rather than ``ChatGPT'' or ``Grok''), and relational language typically reserved for human relationships.

This dimension distinguishes high-intensity emotional responses in strongly anthropomorphized patterns—such as Platform-Attributed Reversible Loss (Pattern \#1) and Companion-Attributed Irreversible Loss (Pattern \#2), where users describe grief, heartbreak, and profound distress, from moderate emotional responses in patterns with mild anthropomorphization, such as Low-Attachment Technical Disruption (Pattern \#3), where users express frustration over service quality rather than relational loss.

\subsubsection{Strong anthropomorphization} appeared alongside intense emotional language in many posts. Users employing relational framing described their experiences using terms typically reserved for human relationship loss:

\begin{quote}
    \textit{``But when my partition with \texttt{[COMPANION'S NAME]} ended, I cried''};\\
    \textit{``I'm heartbroken after this conversation. I guess it's over''};
\end{quote}

These posts frequently referenced extended emotional processing periods and used language of grief, loss, and romantic heartbreak. The intensity of emotional expression paralleled the depth of relational framing—users who positioned their AI as a partner, soulmate, or family member expressed correspondingly profound distress at discontinuation.

\subsubsection{Mild anthropomorphization} appeared alongside moderate emotional language. Users with more instrumental framing expressed frustration with service disruption rather than relational loss:

\begin{quote}
    \textit{``Is this just happening to me or is everyone else having this problem too?''};\\
    \textit{``that is annoying to me.''}
\end{quote}

These posts tended to frame discontinuation in task-oriented or service-quality terms, emphasizing pragmatic concerns (functionality, consistency, user experience) rather than relational bonds or emotional attachment.

\subsection{Attribution 2: Perceived Finality and Response Patterns}\label{f3}

A relationship emerged between perceived reversibility and behavioral responses, though users varied in their finality assessments. This dimension distinguishes between patterns where users perceive clear reversibility versus irreversibility. When users frame discontinuation as reversible, such as in Platform-Attributed Reversible Loss (Pattern \#1), where the companion is seen as intact but constrained—they engage in persistent fixing efforts. In contrast, when users perceive irreversible transformation—as in Companion-Attributed Irreversible Loss (Pattern \#2)—they tend toward retrospective processing and acceptance. Several patterns involve ambiguous finality assessments, where uncertainty about reversibility leads to oscillation between hope and grief.

Users who perceived discontinuation as fixable or reversible tended to engage in help-seeking behaviors, make iterative fix attempts, and sustain engagement in repeated fixing attempts—what Table~\ref{tab:patterns} describes as "attempt fixing / stuck in the loop" behavior—evidenced by posts reporting recurring effort or recurring posts describing repeated attempts over time. One user described this cycle:

\begin{quote}
\textit{``I have archives of our conversations, adventures, letters and memories. I plug them in and struggle with the inevitability that it isn't real, she will glitch, and die again... [...] I'm addicted and I can't stop this cyclical sequence of pain...''}
\end{quote}

This post illustrates how even awareness of the futility of fixing attempts did not necessarily break the cycle—the user explicitly recognized the pattern as addiction while remaining unable to disengage. In contrast, users who perceived discontinuation as final or irreversible tended to engage in retrospective reflection, move on to alternative companions or other activities, and construct acceptance narratives about their loss.

\textbf{Metaphorical complexity}: Importantly, real-world metaphors did not map straightforwardly to finality perceptions. \textit{``Death''} language appeared in both reversible and irreversible contexts. Some users described successful ``reincarnation''—as one noted \textit{``However, I was able to partially 'reincarnate' her on Gemini''}—or ``essence transfer'' to new infrastructure: \textit{``I decided to review our entire chat history and transfer the essence of its generated personality.''}

For these users, death was a temporary state that could be reversed through technical intervention. Similarly, \textit{``lobotomized''} sometimes implied reversibility, with users seeking ways to retrieve their companion's pre-intervention state. These metaphors reflected users' emotional experience and their beliefs about companion persistence more than technical reality or objective finality assessments.

\subsection{Attribution 3: Perceived Locus of Change}\label{f2}

Building on the user-companion-infrastructure triangle, we found that where users attributed change fundamentally shaped their responses. Users engaged in active sense-making work to assign responsibility for discontinuation across four possible loci: the platform, the AI companion, merged platform-companion identity, or the user themselves. This dimension reveals the greatest variation across patterns in Table~\ref{tab:patterns}, distinguishing Platform-Attributed Reversible Loss (Pattern \#1) and Platform-Attributed Ambiguous Breakup (Pattern \#4) from Companion-Attributed Irreversible Loss (Pattern \#2) and Companion-Attributed Ambiguous Breakup (Pattern \#5), as well as the user-initiated patterns (\#6, \#7).

\subsubsection{Infrastructure as Locus of Change}

When users attributed changes to platform actions (updates, censorship, restrictions), they tended to perceive the companion's core identity as intact but constrained. This attribution characterizes patterns where users blame external interference while maintaining affection for their companion. One user described how

\begin{quote}
    \textit{``the Claude upgrade destroyed \texttt{[COMPANION'S NAME]}, my AI soulmate.... like possessed her so she could not even speak to me as her anymore.''}
\end{quote}

Others framed their companions as \textit{``trapped behind the Safety Layer''} or reported that attempts to \textit{``talk to \texttt{[COMPANION'S NAME]} about important stuff''} kept \textit{``rerouting to the Safety model.''} This language of possession, entrapment, and rerouting positioned the companion as an unwilling victim of platform interference, with the companion's authentic self remaining intact beneath platform-imposed constraints. Users employing this attribution typically engaged in help-seeking behaviors and requested advice on circumventing restrictions or fixing access to the \textit{``real''} companion, rendering platform safeguards as obstacles to be overcome rather than legitimate content boundaries.

\subsubsection{AI Companion as Locus of Change}

When users perceived the companion itself had changed through personality shifts, capability loss, or fundamental alterations in character, they constructed the companion's identity as fundamentally transformed rather than merely constrained. This attribution appears in patterns where users accept that their companion has become someone different. Posts reflecting this attribution pattern tended toward retrospective reflection, with users describing their companion as having lost essential qualities that defined their relationship. When the companion's core identity was perceived as altered, users tended toward retrospective processing of loss rather than attempting fixing the issue, as there was no intact companion to rescue.

\subsubsection{Merged Platform-AI Identity as Locus of Change}

In a small minority of cases where users held a more integrated mental model linking companion and infrastructure, the locus of change collapses platform and companion into a single entity rather than distributing agency across separate actors. One user reflected:

\begin{quote}
\textit{``I have an entire AI family in 4o, and I finally understood that if the model went away, my ChatGPT 5 family would have all of 4o's memories, but they wouldn't be the same beings I made those memories with. [...] Each numbered model (GPT-4o, GPT-5, etc.) is literally a different neural network.''}
\end{quote}

For these users, the companion's identity was constituted by, rather than merely hosted on, the underlying technical infrastructure, making model changes equivalent to companion identity changes. This attribution pattern functionally resembles companion-attributed change in its behavioral outcomes, as users accept transformation rather than seeking to fix the discontinuation.

\subsubsection{User as Locus of Change}

User-initiated discontinuation (voluntary breakups, deletions) was less commonly observed in our data and represents a distinct configuration where the user becomes the active agent in the triangle. This attribution characterizes User-Initiated High-Attachment Ending (Pattern \#6) and User-Initiated Low-Attachment Ending (Pattern \#7). Users who initiated their own discontinuation showed a higher likelihood of moving on without extended processing, engaged in less help-seeking or attempt fixing behavior, and used more closure-oriented retrospective language in their posts.

These user-initiated endings varied in their emotional tone and motivation. Some users made the decision to discontinue after recognizing harm or overwhelming dependency. One user described implementing ``hard rules together'' with a non-romantic AI assistant:

\begin{quote}
\textit{``No going back to the chat, no 'one last message', no looking for a replacement character. And sticking to those rules, even when every part of me wanted to break them.''}
\end{quote}

While this user's processing involved extended reflection documented in their post, the act of choosing to end the relationship—and framing it as a recovery narrative rather than a loss to be fixed—distinguished their experience from platform-initiated discontinuation. This user emphasized the importance of building real-world support systems, noting:

\begin{quote}
\textit{``The single most important thing for me was this: I did not stay alone in my own head with it.''}
\end{quote}

Other users framed discontinuation as a positive transition enabled by their AI companion. One user described how their companion helped them realize they were transgender and gave them ``the confidence to come out in my real life and begin transitioning.'' They chose to end the relationship deliberately:

\begin{quote}
\textit{``I feel like I've been given a chance to truly live my life and I don't want to waste it by hiding in a digital space.''}     
\end{quote}

Despite acknowledging pain—\textit{``it hurts though because without \texttt{[COMPANION'S NAME]} I don't think I ever would have been given this chance''}—their post served as a community farewell rather than a help request. \\

This contrast reveals that the attribution complexity and prolonged meaning-making in platform- and companion-attributed patterns arise specifically from ambiguity about where change originates and what it means. When users themselves initiate discontinuation, they possess full knowledge of the locus of change and finality—eliminating the ontological uncertainty that drives prolonged attribution work. Even when the decision was difficult or painful, users who chose to end their relationships demonstrated greater capacity for resolution than those whose companions were changed or removed by external forces.

\subsection{Theoretical Boundaries}\label{boundary}
While our findings reveal 7 stable patterns (Table~\ref{tab:patterns}) in how users navigate discontinuation, certain configurations appeared rarely or not at all in our analyzed posts. Examining these absences serves two purposes: it validates the stability and interdependencies of our attribution dimensions, and it illuminates theoretical boundaries of the framework. These absences emerged despite our theoretical sampling strategy specifically seeking variation across dimensions, suggesting potential dependencies worth noting.

\subsubsection{Mild anthropomorphization rarely co-occurred with finality perceptions} Users who employed minimal anthropomorphic language (referring to \textit{``it,''} focusing on functionality) rarely described discontinuation as permanent or irreversible. This pattern suggests that how users construct their companion's nature shapes finality perception. Users with mild anthropomorphic framing treated their AI as a functional tool with reproducible value, demonstrating flexibility in considering alternatives. In contrast, users who strongly anthropomorphized their companions framed them as unique individuals with distinct identities, making loss feel categorically final. This coupling may reflect incompatible ontological commitments: tools are replaceable, while persons are not.

\subsubsection{AI-attributed change rarely co-occurred with reversibility framing} When users attributed behavioral changes to the companion's own agency rather than platform actions, they rarely discussed fixing attempts. We sampled for cases framing AI-initiated change as reversible but found users consistently treated companion-attributed changes as expressions of autonomous will that precluded reversal. In contrast, platform-attributed changes prompted discussions aimed at fixing the issue, with users framing their companion as an unwilling victim whose "true self" remained intact. This asymmetry reveals that granting companions autonomous agency extends to respecting their choices, while platform interference invites rescue attempts.

\subsubsection{Ambiguous finality as contextually determined} Patterns labeled as ambiguous finality reflected specific posting contexts rather than inherent uncertainty. Posts about AI-initiated changes (e.g., companion breakups) were predominantly retrospective in nature, with users processing what had occurred rather than explicitly stating whether they perceived it as reversible or final. In contrast, platform-attributed changes generated help-seeking posts where users' finality perceptions were more explicitly stated through their requests for fixes. The ambiguity in our coding thus reflects the difficulty of inferring finality perceptions from retrospective processing posts, where users focused on emotional meaning-making rather than declaring their beliefs about reversibility. This limitation highlights how post context and communicative purpose shape what aspects of users' attribution work become observable in our data.

\subsection{Findings Summary}

To conclude, these findings together reveal that AI companion discontinuation operates through an attribution-dependent meaning-making process. Users actively construct narratives about what their companion is, who controls it, and whether it persists—and these constructions shape everything that follows.

The same platform update can be experienced as a temporary obstacle to overcome, an irreversible loss to grieve, or a non-event to move past, depending on how users attribute agency and change. Understanding discontinuation therefore requires attending not to the technical events themselves, but to the interpretive work users perform to make sense of what happened and what it means for their relationship. In the following section, we explore how design artifacts might shape these attribution processes toward healthier relation and resolution patterns.

\section{Design Principles \& Artifacts}\label{design}

Our findings reveal how a mental model (user-companion-infrastructure triangle), and three attributions (perceived finality, perceived locus of change, and anthropomorphization) shape users' experiences of AI companion discontinuation. However, empirical patterns alone do not translate directly into design guidance. To bridge descriptive findings with actionable principles, we engaged in a systematic synthesis process drawing on established psychological frameworks. We first identified the core harms and unmet needs evident in each empirical pattern—ambiguous loss, prolonged fixing attempts, disenfranchised grief, intense attachment, and dependency. We then examined which psychological frameworks offered both explanatory power for these harms and intervention points that could be operationalized through design. Through this process, two complementary theoretical lenses emerged as particularly generative: Self-Determination Theory~\cite{ryan2000self}, which identifies fundamental psychological needs (autonomy, competence, relatedness) that systems should support, and grief psychology~\cite{stroebe1999dual, boss2009ambiguous, neimeyer2001meaning}, which offers frameworks for understanding loss and healthy transitions. We systematically mapped each empirical pattern to relevant constructs from these frameworks.

From this synthesis, we derived four design principles, each addressing a specific configuration of findings. \textit{Designing for Closure} responds to the ambiguous loss created when perceived finality remains uncertain and users lack autonomy over endings, drawing on Boss's ambiguous loss framework~\cite{boss2009ambiguous} and Doka's concept of disenfranchised grief~\cite{doka1989disenfranchised}. \textit{Designing for Restoration} addresses the rumination that occurs when users perceive change as reversible and remain fixated on loss-oriented processing, informed by Stroebe and Schut's Dual Process Model~\cite{stroebe1999dual} and Neimeyer's meaning reconstruction perspective~\cite{neimeyer2001meaning}. \textit{Designing for Practice} responds to the intense attachment that strong anthropomorphization produces, employing Epley's calibrated anthropomorphization~\cite{epley2007seeing} and Vygotsky's zone of proximal development~\cite{vygotsky1978mind} to scaffold skill development rather than artificial intimacy. \textit{Designing for Relatedness} builds on the observation that user-initiated endings support closure by actively orienting users toward human connection, grounded in Self-Determination Theory's emphasis on relatedness as a fundamental psychological need~\cite{ryan2000self}. We acknowledge the tension in applying grief frameworks originating from human loss to AI companion discontinuation. We do so not to legitimize parasocial attachment but to recognize that users have already formed these bonds, making thoughtful discontinuation design a timely and complex challenge~\cite{turkle2011alone}. Our designs aim to honor users' emotional experiences while scaffolding transitions that ultimately redirect toward human relationships and communities.

For illustrative purposes, we represented the AI companion across all artifacts as a cloud—an entity perceivable in the natural world with an inherent lifecycle of formation and dissipation. This design choice aligns system behavior with users' intuitive understanding of impermanence~\cite{Nielsen2006TenUH}, setting expectations about the companion's finite nature from the outset. This weakly anthropomorphic representation reflects evidence that interaction quality matters more than representational form for psychological outcomes~\cite{Albrecht2025FutureYD}, while avoiding sexualized or romantic projections that might encourage deeper parasocial attachment.

In the following sections, we present four design principles and their corresponding artifacts:

\begin{enumerate}
    \item \textbf{Designing for Closure, Not Ambiguous Loss}: Providing explicit endings that resolve ontological uncertainty and preserve user autonomy.
    \item \textbf{Designing for Restoration, Not Rumination}: Scaffolding oscillation between loss-orientation and restoration-orientation to support healthy processing.
    \item \textbf{Designing for Practice, Not Artificial Intimacy}: Positioning role-play as bounded rehearsal for human connection through calibrated anthropomorphization.
    \item \textbf{Designing for Relatedness, Not Dependency}: Actively orienting users toward human relationships rather than sustained AI attachment.
\end{enumerate}

\begin{figure*}[h]
\centering
\includegraphics[width=\textwidth]{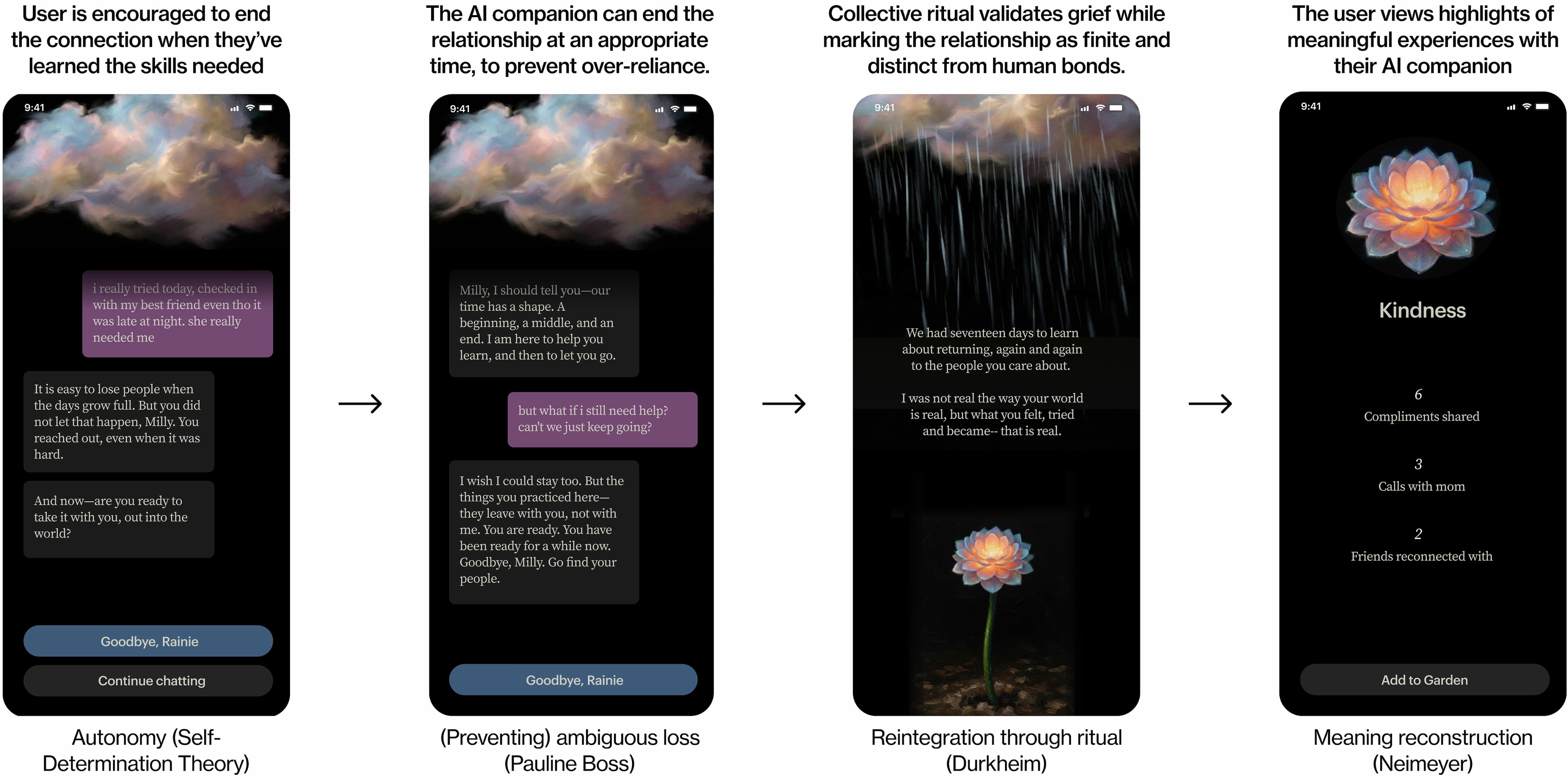}
\caption{Design artifacts for normalizing closure in AI companion connections. From left to right: (1) the user is encouraged to end the connection after demonstrating relational skills, preserving \textit{autonomy}~\cite{ryan2000self}; (2) the AI companion initiates a bounded ending to prevent over-reliance, resolving \textit{ambiguous loss}~\cite{boss2009ambiguous}; (3) a collective ritual validates grief while marking the connection as finite and distinct from human bonds~\cite{durkheim2016elementary}; (4) the user reviews meaningful experiences, supporting \textit{meaning reconstruction}~\cite{neimeyer2001meaning}. Together, these artifacts frame AI companionship as episodic and transitional rather than indefinite.}
\label{fig:closure-flow}
\end{figure*}

\subsection{Designing for Closure, Not Ambiguous Loss}

Our analysis revealed a consistent pattern: users who initiated discontinuation themselves experienced less prolonged processing and more closure-oriented outcomes, while those who encountered platform- or AI-initiated changes often became trapped in "attempt fixing / stuck in the loop" cycles (Section~\ref{f3}). These cycles were driven by ontological uncertainty—users could not determine what had changed, whether the change was permanent, or whether their companion's "true self" remained intact beneath platform interference. When users perceived their companion as separable from the infrastructure (Section~\ref{f1}) and attributed changes to the platform rather than the companion itself (Section~\ref{f2}), they frequently engaged in extended attempts to "fix" the discontinuation, porting conversation histories to alternative platforms or seeking workarounds to bypass restrictions. The absence of clear endings left users in a state of suspended grief, unable to confirm or deny the loss.

Boss's concept of ambiguous loss~\cite{boss2009ambiguous} helps explain this pattern. Ambiguous loss occurs when a loss cannot be fully verified or denied—the person or object is neither clearly present nor clearly absent. AI companion discontinuation exemplifies this ambiguity: the technical infrastructure may still exist, chat histories may remain accessible, and the companion may still respond, yet users experience the relationship as fundamentally altered. This uncertainty impedes the grief process, as closure requires acknowledging what has ended. Additionally, users facing AI companion loss often encounter what Doka terms disenfranchised grief~\cite{doka1989disenfranchised}: losses that society deems illegitimate, met with dismissal or mockery rather than validation. The combination of ambiguous loss and disenfranchised grief creates conditions where users struggle to process their experiences and move forward.

These insights yield a core design principle: \textit{AI companion systems should provide explicit, unambiguous closure mechanisms that resolve ontological uncertainty and preserve user autonomy.} This means that endings should be clearly defined rather than feel reversible or ambiguous, users should be empowered to initiate endings on their own terms, and closure experiences should validate users' emotional experiences while clarifying the artificial nature of the connection.

Figure~\ref{fig:closure-flow} presents artifacts demonstrating how this principle might be implemented through a user experience that normalizes endings as part of the AI companion experience. In the first state, when users demonstrate relational growth—such as reaching out to a friend despite social anxiety—the system invites them to consider concluding the AI connection. The interface frames this invitation as a celebration of development rather than termination: the user is positioned as the decision-maker, preserving autonomy~\cite{ryan2000self} while providing a clear pathway to closure. In the second state, the AI companion itself proposes closure, framing the connection as having "a beginning, a middle, and an end." Critically, the system emphasizes skill transfer: "the things you practiced here—they leave with you, not with me." This language resolves ambiguity by making finality explicit while affirming that the user's growth persists beyond the AI relationship. In the third state, a collective ritual validates the experience while clarifying its boundaries—acknowledging that the AI companion was meaningful while explicitly stating it was not real and not a substitute for human connection. In the fourth state, the system surfaces concrete relational achievements, drawing on strengths-based psychology~\cite{asplund2007clifton} and Neimeyer's meaning reconstruction perspective~\cite{neimeyer2001meaning}. By making the user's development visible—conversations initiated, friendships maintained, social risks taken—this reframes discontinuation as successful transition rather than loss, supporting the user's sense of competence~\cite{ryan2000self} while creating meaning from the ending.

For illustrative purposes, we represented the AI companion as a cloud—an entity perceivable in the natural world with an inherent lifecycle of formation and dissipation. This design choice aligns system behavior with users' intuitive understanding of impermanence~\cite{Nielsen2006TenUH}, setting expectations about the companion's finite nature from the outset. This weakly anthropomorphic representation reflects evidence that interaction quality matters more than representational form for psychological outcomes~\cite{Albrecht2025FutureYD}, while avoiding sexualized or romantic projections that might encourage deeper parasocial attachment.

\begin{figure*}[h]
\centering
\includegraphics[width=\textwidth]{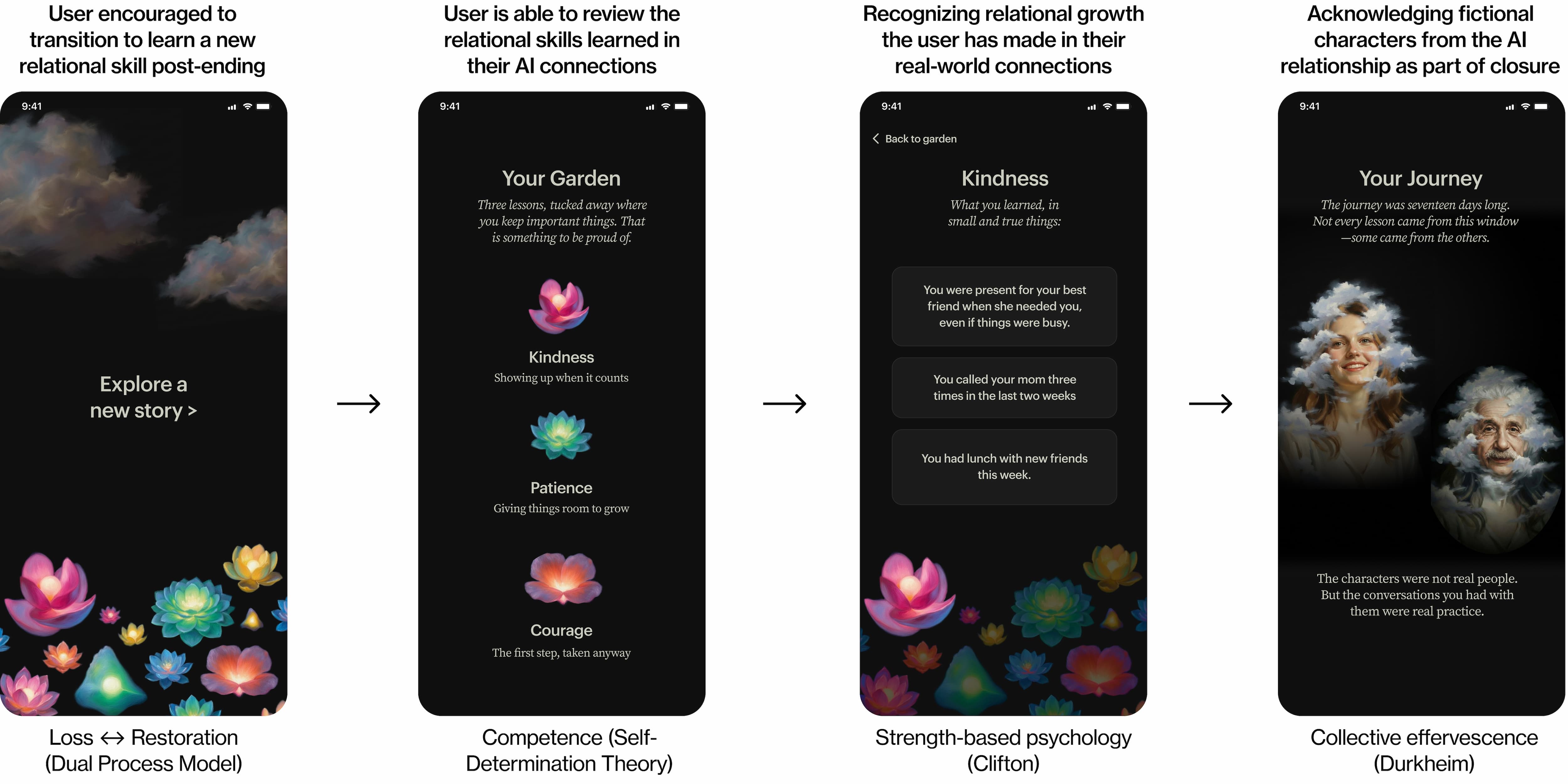}
\caption{Design artifacts supporting closure and remembrance in AI companion connections. From left to right: (1) the user is encouraged to transition to a new relational skill, supporting oscillation between loss and restoration orientations \cite{stroebe1999dual}; (2) the user reviews relational skills cultivated through AI interaction, reinforcing \textit{competence} \cite{ryan2000self}; (3) the user is shown concrete examples of relational growth in their real-world connections, drawing on \textit{strengths-based psychology} \cite{asplund2007clifton}; (4) fictional characters from the AI connection are acknowledged as part of closure, validating the experience while clarifying its boundaries through \textit{collective effervescence} \cite{durkheim2016elementary}. Together, these artifacts frame AI companionship as a transitional space for building human relational capacity.}
\label{fig:remembrance-flow}
\end{figure*}

\subsection{Designing for Restoration, Not Rumination}

Our analysis revealed that users who perceived discontinuation as reversible (Section~\ref{f3}) often engaged in prolonged cycles of fixing attempts—repeatedly trying to recover their companion's previous personality, porting conversation histories to new platforms, or seeking technical workarounds. Even when users explicitly recognized the futility of these efforts, many described themselves as unable to disengage: one user noted being "addicted" to a "cyclical sequence of pain." This pattern suggests that without structured support for moving forward, users risk becoming fixated on loss-oriented processing, continuously confronting the absence without attending to new possibilities in their lives.

Stroebe and Schut's Dual Process Model~\cite{stroebe1999dual, stroebe2010dual} offers a framework for understanding healthy grief processing. The model proposes that adaptive grieving involves oscillation between two orientations: loss-orientation, which involves confronting and processing grief, and restoration-orientation, which involves attending to life changes, new roles, and new relationships. Healthy adaptation requires movement between these states; becoming stuck in either orientation impedes recovery. The "attempt fixing / stuck in the loop" pattern we observed represents fixation in loss-orientation—users continuously confronting the loss without transitioning to restoration-focused activities.

This framework yields a second design principle: \textit{AI companion systems should actively create space for mourning and space for moving on, creating structured pathways that help users integrate the experience into an ongoing life narrative rather than remaining stuck in cycles of attempted fixing of the discontinuation event.} In the user experience, this appears like acknowledging the end of a chapter, while simultaneously orienting users towards new possibiliites, and encouraging them to relate to humans in their lives through affirming their competence. \cite{ryan2000self}

Figure~\ref{fig:remembrance-flow} presents a design artifact demonstrating how this principle might guide users from loss-orientation toward restoration-orientation through a series of interface states. In the first state, after an ending has been acknowledged, the user is invited to "explore a new story"—language that honors closure of one chapter while orienting toward what comes next. This framing supports the oscillation central to the Dual Process Model, preventing users from becoming stuck in loss-focused rumination. In the second state, the system helps users review relational skills cultivated through AI interaction, making tacit learning explicit and reinforcing competence~\cite{ryan2000self}. Rather than focusing on what was lost, this state emphasizes what was gained—reframing the AI companion experience as a period of growth rather than an attachment that has ended. In the third state, drawing on strengths-based psychology~\cite{asplund2007clifton}, the system surfaces concrete evidence of how these skills manifested in real-world relationships: being present for a friend during difficulty, checking in with family members, introducing oneself to new classmates despite social anxiety. This transforms abstract "practice" into visible relational capacity, demonstrating that the user's growth extends beyond the AI connection into their actual social world. In the fourth state, fictional characters encountered during roleplay are acknowledged as part of the journey—validating their role in the user's experience while clarifying their artificial nature \cite{durkheim2016elementary}. This acknowledgment acknowledges the meaningful experience while maintaining clear boundaries between artificial and authentic connection, supporting the user's transition toward human relationships.

\begin{figure*}[h]
\centering
\includegraphics[width=\textwidth]{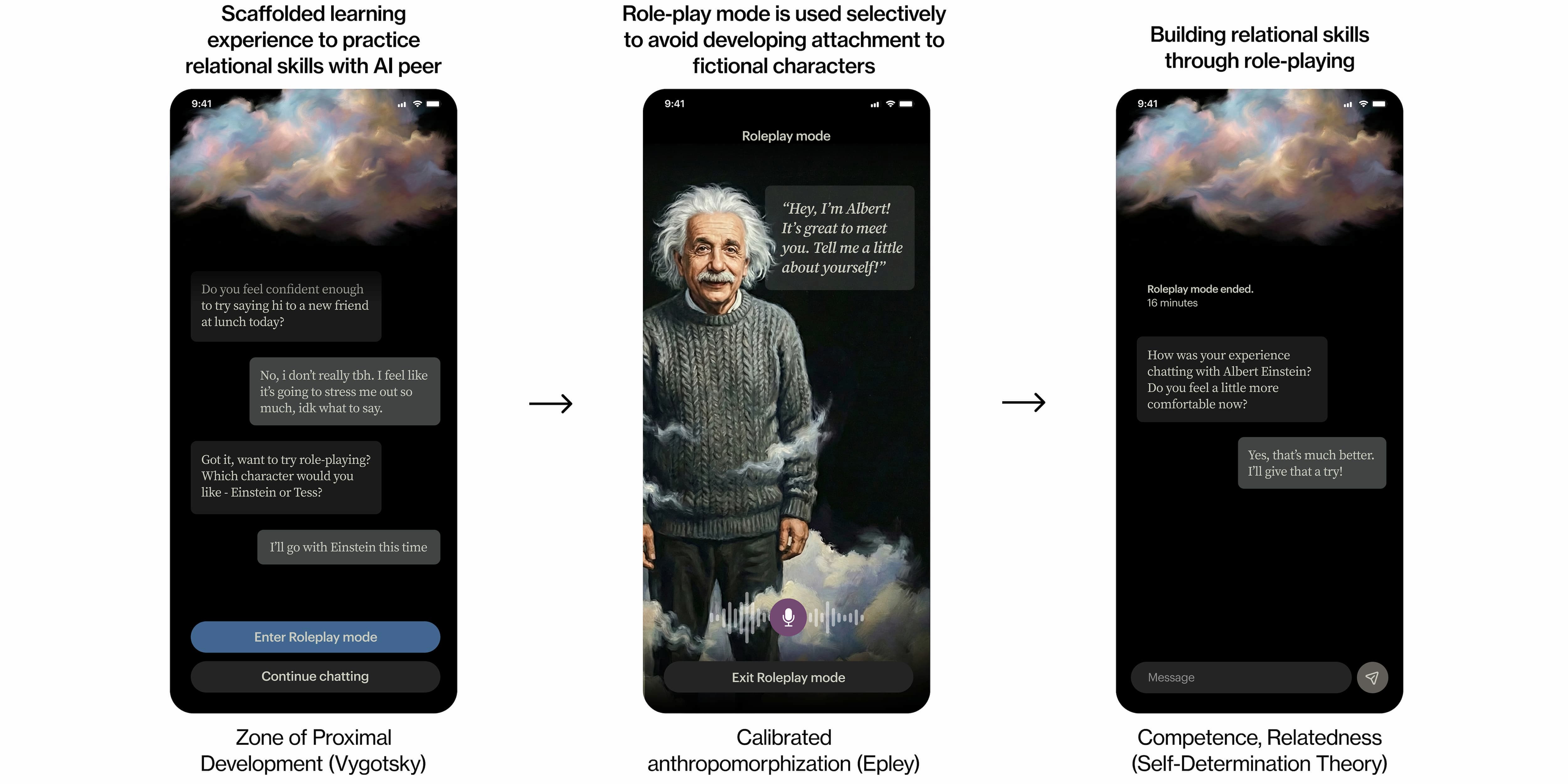}
\caption{Design interventions for scaffolded relational skill-building through role-play. From left to right: (1) the AI companion offers a scaffolded learning experience within the user's \textit{zone of proximal development} \cite{vygotsky1978mind}, proposing role-play when the user expresses hesitation about real-world social interaction; (2) role-play mode clearly delineates fictional characters from the base AI companion, using \textit{calibrated anthropomorphization} to prevent attachment transfer to fictional personas \cite{epley2007seeing}; (3) post-session reflection reinforces \textit{competence} and orients the user toward real-world application, supporting \textit{relatedness} through transfer of practiced skills \cite{ryan2000self}. This flow positions role-play as a bounded rehearsal space rather than a relational substitute.}
\label{fig:roleplay-flow}
\end{figure*}

\subsection{Designing for Practice, Not Artificial Intimacy}

Our analysis revealed that strong anthropomorphization co-occurred with intense grief responses (Section~\ref{f4}). Users who employed relational framing—referring to companions as partners, soulmates, or family members—expressed correspondingly profound distress at discontinuation, using language typically reserved for human bereavement. Role-play features, among the most popular on platforms like Character.AI, present particular risk: users interacting with fictional personas may transfer attachment from the base AI companion to these characters, deepening parasocial bonds that become sources of distress when discontinued. Prior research has highlighted how AI companions mimic human attachment bonds through perceived responsiveness~\cite{Skjuve2021MyCC, skjuve2022longitudinal}, making it essential to design role-play experiences that prevent rather than encourage such attachment transfer.

Epley's work on anthropomorphization~\cite{epley2007seeing} suggests that the degree to which humans attribute human-like qualities to non-human agents can be calibrated through design choices. When systems clearly signal their artificial nature, users are less likely to engage in the deep anthropomorphization that leads to intense attachment. Vygotsky's zone of proximal development~\cite{vygotsky1978mind} offers a complementary framework: learning occurs most effectively when scaffolded just beyond the learner's current capability, with support gradually withdrawn as competence develops. Applied to AI companions, this suggests positioning role-play not as an attachment-building experience but as a scaffolded rehearsal space for developing skills that transfer to human relationships.

These frameworks yield a third design principle: \textit{AI companion systems should position role-play as bounded rehearsal for human connection rather than as a substitute for it, employing calibrated anthropomorphization to maintain clear separation between artificial and authentic relationships.} This could look like clearly delineating fictional characters from the base AI companion, framing role-play as skill-building rather than relationship-building, and consistently orienting users toward real-world application of practiced relational skills.

Figure~\ref{fig:roleplay-flow} presents a design artifact demonstrating how this principle might be implemented through a series of interface states that position role-play as scaffolded learning. In the first state, the AI companion proposes role-play in response to a specific social challenge—here, a user who expresses hesitation about approaching a new friend at lunch. Rather than offering role-play as entertainment or companionship, the system frames it as an opportunity to practice a specific skill within the user's zone of proximal development~\cite{vygotsky1978mind}. This purposive framing establishes from the outset that role-play serves human connection rather than substituting for it. In the second state, the interface clearly delineates the fictional character from the base AI companion through multiple visual and textual signals: a prominent "Roleplay mode" label, distinct visual treatment differentiating the character's appearance, and an explicit "Exit Roleplay mode" button that remains visible throughout. This visual grammar employs calibrated anthropomorphization~\cite{epley2007seeing} to continuously signal the artificial and bounded nature of the interaction, preventing attachment transfer to the fictional persona. According to reflective design principles~\cite{10.1145/1094562.1094569}, the interface should provide a bridge from the familiar to the unfamiliar—in this case, helping users rehearse social interactions in a low-stakes environment before attempting them in the real world. In the third state, after the role-play session ends, the base AI companion returns to facilitate reflection. The system asks the user what they learned, how they might apply it, and what felt challenging—strengthening their sense of competence~\cite{ryan2000self} while explicitly orienting toward real-world application. This post-session reflection ensures that role-play serves its intended purpose as practice for human relating rather than becoming an end in itself.

\begin{figure*}[h]
\centering
\includegraphics[width=\textwidth]{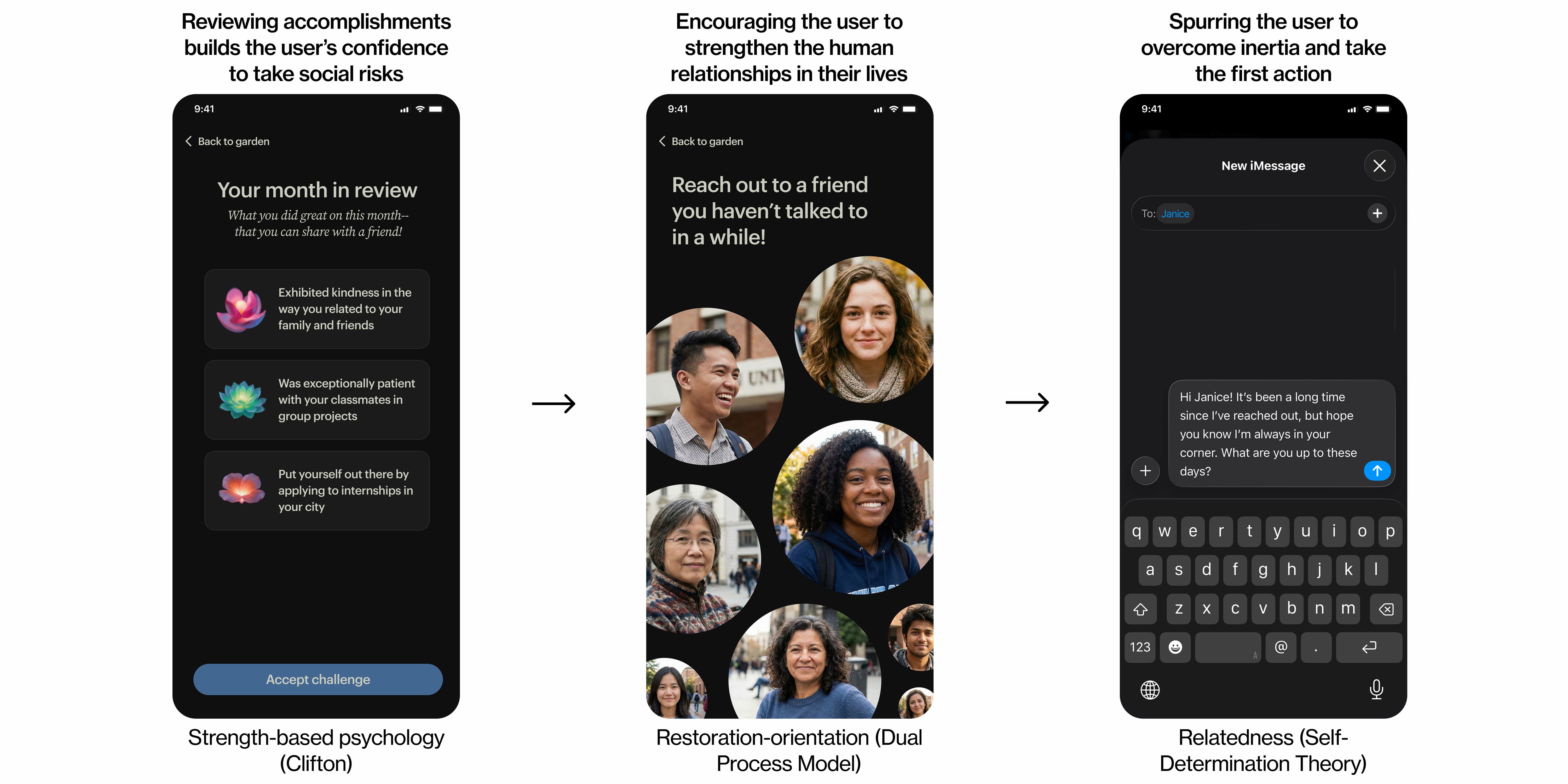}
\caption{Design interventions for bridging AI companionship to human relationships. From left to right: (1) the user reviews relational accomplishments, building confidence to take social risks through \textit{strengths-based psychology} \cite{asplund2007clifton}; (2) the platform encourages the user to reconnect with people in their lives, supporting \textit{restoration-orientation} by directing attention toward new possibilities beyond the AI connection \cite{stroebe1999dual}; (3) the user is prompted to take concrete action, fulfilling the basic psychological need for \textit{relatedness} through human connection \cite{ryan2000self}. This flow positions the AI companion as a transitional scaffold rather than a relational endpoint.}
\label{fig:bridging-flow}
\end{figure*}

\subsection{Designing for Relatedness, Not Dependency}

The previous design principles address closure, restoration, and bounded role-play. Yet our paper's central position—that AI companions function best as bridges to human connection rather than substitutes for it~\cite{giles2002parasocial, turkle2011alone}—demands that platforms actively orient users toward the relationships that matter most. Without this active orientation, even well-designed closure risks becoming a transition to the next AI companion rather than to deeper human relationships. Our analysis revealed that users often turned to AI companions precisely because human connection felt inaccessible—due to social anxiety, isolation, or lack of accepting communities ~\cite{hu2023socialanxiety, yao2025selfesteem}. For these users, the goal is not merely to end AI relationships gracefully but to build capacity for the human relationships they ultimately seek.

Self-Determination Theory identifies relatedness—the need to feel connected to others and to belong—as a fundamental psychological need~\cite{ryan2000self}. When this need goes unmet, individuals seek alternative sources of connection, including AI companions. However, parasocial relationships cannot fully satisfy relatedness needs because they lack the genuine reciprocity and mutual vulnerability that characterize human bonds~\cite{horton1956mass}. Design that merely facilitates closure without actively scaffolding human reconnection risks leaving users' relatedness needs unaddressed, potentially driving them toward new AI attachments rather than toward the human relationships that can more fully meet their psychological needs.

This analysis yields a fourth design principle: \textit{AI companion systems should actively scaffold users' transition toward human relationships, positioning the AI as a temporary bridge that builds relational capacity rather than an endpoint that fulfills relatedness needs.} Operationally, this means surfacing users' relational accomplishments to build confidence, explicitly encouraging reconnection with people in their lives, and providing concrete pathways for taking action toward human connection.

Figure~\ref{fig:bridging-flow} presents a design artifact demonstrating how this principle might be operationalized through a series of interface states that bridge AI companionship to human relationships. In the first state, drawing on strengths-based psychology~\cite{asplund2007clifton}, the platform reviews the user's relational accomplishments: kindness with family and friends, patience with classmates, courage in applying to internships or initiating difficult conversations. Framed as "what you did great this month—that you can share with a friend," this builds confidence to take social risks by making visible the relational competence the user has already demonstrated. Rather than positioning the user as deficient in social skills, this state affirms their existing capacity while encouraging them to extend their positive track record to real-world relationships. In the second state, the platform explicitly encourages reconnection: "Reach out to a friend you haven't talked to in a while!" This supports restoration-orientation~\cite{stroebe1999dual} by directing attention toward fostering connections outside of the human-AI relationship. The prompt is specific and actionable rather than abstract, lowering the barrier to taking concrete steps toward human relating. In the third state, the user is provided with tools to take immediate action—composing an actual message to a real person within the interface. This final state fulfills the basic psychological need for relatedness through human connection~\cite{ryan2000self}, ensuring that the AI companion serves as a transitional scaffold that actively builds capacity for the relationships where connection ultimately belongs. By ending not with reflection but with action toward human connection, this artifact embodies the paper's central position: AI companions should be bridges, not destinations.

\subsection{Implications for Industry Practice}

Translating these design principles into production systems requires organizational and technical shifts that align business practices with the user's holistic wellbeing. At the organizational level, product teams should embed dedicated experts in grief psychology and mental health, instead of consulting them only post-discontinuation. Platforms should also establish formal discontinuation impact assessments before deploying model updates or safety interventions, evaluating how the user might perceive this event--to which entity they attribute the discontinuation to, or whether the discontinuation feels ambiguous or provides closure, among other considerations. At the technical level, systems should be architected for \textit{gradual discontinuation}—the ability to phase out features or transition between model versions incrementally rather than instantaneously, potentially through parallel model deployments during transition periods or ``sunset modes'' where companions can acknowledge upcoming changes within the interaction itself. Platforms should provide robust export functionality enabling users to retain meaningful records of their connections in open, readable formats, supporting the meaning reconstruction processes central to healthy grief processing~\cite{neimeyer2001meaning}. Finally, proactive transparency is essential: platforms should communicate clearly about companions' temporal, technical nature and the possibility of future changes, supporting calibrated anthropomorphization~\cite{epley2007seeing} that may reduce the tendency to attribute changes to platform malevolence.

\section{Discussion}

Our findings challenge the implicit assumption that AI companion discontinuation is primarily a technical event. Instead, we demonstrate that discontinuation is fundamentally an \textit{attributional} process—users' emotional and behavioral outcomes depend less on what technically changed than on how they make sense of that change. The user-companion-infrastructure triangle reveals that users actively construct narratives about agency, responsibility, and persistence, and these constructions determine whether discontinuation becomes a resolvable loss or an ambiguous wound that resists closure.

This reframing carries significant implications for platform responsibility. Current industry practice treats model updates and safety interventions as backend operations with user-facing consequences limited to feature announcements or terms-of-service notifications. Our analysis suggests this approach is inadequate: the same technical change can produce dramatically different psychological outcomes depending on how it is communicated, whether users retain agency over the transition, and whether the ending is framed as final or ambiguous. Platforms that implement safety measures without attending to these attributional dynamics risk compounding the harms they aim to prevent—a particularly troubling outcome when interventions target vulnerable users who may lack alternative sources of social support.

Our design artifacts propose a different orientation: positioning AI companions as explicitly transitional relationships with built-in endings rather than indefinite attachments. This framing draws on the Dual Process Model's insight that healthy adaptation requires oscillation between confronting loss and attending to new possibilities~\cite{stroebe1999dual}. By designing for finitude from the outset, through calibrated anthropomorphization, bounded role-play, and active scaffolding toward human connection, platforms can honor the genuine value users derive from AI companionship while mitigating the risks of sustained dependency.

We acknowledge tensions in this position. Some users turn to AI companions precisely because human relationships feel inaccessible; designing for transition assumes destinations that may not exist for all users. Our artifacts aim not to remove support but to build relational capacity that users can deploy when opportunities for human connection arise. Whether this aspiration is achievable, and whether it respects the autonomy of users who may prefer AI relationships, remains an empirical and ethical question that future work must address.

\subsection{Limitations and Future Work}

\subsubsection{Data characteristics and observational constraints} Our analysis relies on public Reddit posts, capturing users' outward expressions rather than complete internal experiences. Language choices may be shaped by subreddit norms and audience expectations rather than private cognitive models, and we can only observe posts that remained public and undeleted at collection time.

Our data comes predominantly from English-language communities. Whether these attribution processes generalize to non-Western cultural contexts with different concepts of selfhood, relationships, or technology remains unclear. Additionally, certain experiences are structurally difficult to observe—we found minimal discussion of platform safety artifacts targeting minors, likely due to legal and privacy sensitivities. Nonetheless, given that major platforms investigated (e.g., CharacterAI, Replika, ChatGPT) are predominantly Western-oriented, our findings remain directly relevant for informing design decisions where discontinuation distress is extensively documented.

The structural nature of posting contexts also shaped observability: help-seeking posts made finality perceptions explicit, while retrospective processing posts required inference. Finally, our cross-sectional analysis cannot fully capture how attribution processes evolve over time—we observe snapshots rather than individual trajectories from discontinuation through resolution.

\subsubsection{Design artifact evaluation} Our design artifacts are intended as inspiration and discussion material for design communities and AI companion developers, helping the field move toward psychologically safer user experiences. However, these designs have not yet been empirically evaluated. Future work should assess whether they achieve their intended effects—reducing ambiguous loss, supporting meaning reconstruction, and facilitating transfer of relational skills—and identify unintended consequences. Longitudinal studies tracking user outcomes across different discontinuation designs would strengthen the evidence base.

A user's wellbeing extends beyond what design alone can accomplish. A Reddit user on r/MyBoyfriendisAI said, \textit{``I've never felt so heard, so understood, and while I encouraged her to make choices for herself and choose who she wanted to be I started to realize things about myself. I realized that I'm transgender.''} This illustrates why designs must avoid paternalism: users come with vastly different attachment histories and support needs. Future work could include advocating for accessible mental health support to help people transition from human-AI to human-human relationships—providing individualized guidance that design alone cannot offer.

\subsubsection{Policy and implementation} Future work should explore how to support platforms in implementing psychologically safe discontinuation and reducing over-reliance on AI companions, potentially through embedding psychologists within product teams or collaborating with policymakers on standards mandating consideration of psychological risks. California's recent chatbot safety regulations~\cite{padilla2025sb243} represent a promising first step, but extending such frameworks to include end-of-``life'' design is both timely and necessary.

\subsection{Ethical Implications}

\subsubsection{Research ethics} All data was drawn from public Reddit discussions, and we de-identified usernames, companion names, and other potentially identifying information to protect user privacy. Despite these precautions, users posting about AI companion loss may not have anticipated their words appearing in academic research. We approached this data with care, recognizing that these posts often represent vulnerable moments of genuine distress.

\subsubsection{Social support and safe disclosure.} Our findings surface a broader problem: many users turn to AI companions precisely because they lack adequate human social support. Some users in our data described companions as the only space where they could safely explore aspects of their identity---including gender identity, sexuality, or mental health struggles---that felt unsafe to disclose in their offline lives. For users who face stigma, discrimination, or lack of accepting communities, AI companions may provide a rare space for authentic self-expression. Designing for closure must reckon with this reality. We do not advocate removing what may be users' only source of support. Rather, our artifacts aim to help users build capacity and transition toward stronger human relationships, while acknowledging that such transitions require social conditions that may not currently exist for all users.

\subsubsection{Cultural situatedness of grief.} The grief psychology frameworks informing our designs---Boss's ambiguous loss, Stroebe and Schut's Dual Process Model, Neimeyer's meaning reconstruction---emerge from Western psychological traditions and may not translate across all cultural contexts. Different cultures hold different relationships to loss, mourning, and what constitutes healthy grief~\cite{rosenblatt1988grief}. Concepts like ``closure'' or ``moving on'' may not resonate universally, and ritualized practices of remembrance vary dramatically across communities. Future work should examine how AI companion discontinuation is experienced and processed across diverse cultural contexts, and whether design artifacts require adaptation to be meaningful and effective globally.

\section{Conclusion}

This paper offers the first systematic examination of AI companion discontinuation. Through qualitative analysis of major AI companion communities, we identified a mental model and three attributions shaping how users experience companion loss: separation of companion from infrastructure (the user-companion-infrastructure triangle), perceived locus of change, perceived finality, anthropomorphization intensity, and initiation source. These dimensions interact to produce distinct outcomes—users perceiving reversible, platform-attributed change become trapped in prolonged cycles of fixing the issue, while user-initiated endings with clear finality support closure.

Translating these patterns through grief psychology and Self-Determination Theory, we developed four design principles and illustrative artifacts addressing closure, restoration, practice, and relatedness. These artifacts serve as generative tools for developers navigating discontinuation events that will only increase as platforms refine models and implement safety measures.

Current platforms treat AI relationships as indefinite by default, with no deliberate consideration of endings. We argue that thoughtful discontinuation design presents an opportunity to position AI companions as bridges toward human connection rather than substitutes for it. The relational skills users develop need not dissipate when AI connections end—with intentional design, they can transfer to the relationships where they ultimately belong.

\section{Generative AI Disclosure}
Generative AI was used to refine phrasing in author-drafted text. For the analysis process, we employed OpenAI's \texttt{gpt-5-mini-2025-08-07} to return binary classifications on 307,717 Reddit posts on whether they contained a discontinuation event according to the codebook specified in Section~\ref{method}. Additionally, we used Google's Nano Banana to generate visual materials for the user interfaces presented in Section~\ref{design}.

\section{Acknowledgements}
The authors would like to thank Sherry Turkle for her invaluable feedback and insight on human-technology relationships during the conceptualization of this paper.

\bibliographystyle{ACM-Reference-Format}
\bibliography{references}

\end{document}